\documentclass[aps,superscriptaddress,amsfonts,amssymb,
amsmath,showpacs,twocolumn]{revtex4}
\date{\today}

\begin{document}

\author{G.\ E.\ A.\ Matsas}\email{matsas@ift.unesp.br}
\address{Instituto de F\'\i sica Te\'orica, Universidade 
Estadual Paulista,
Rua Pamplona 145, 01405-900, S\~ao Paulo, SP, Brazil}
\author{M.\ Richartz}\email{richartz@ifi.unicamp.br}
\address{Instituto de F\'\i sica Gleb Wataghin, UNICAMP, C.\ P.\
6165, 13083-970, Campinas, SP, Brazil}
\author{A.\ Saa}\email{asaa@ime.unicamp.br}
\address{Departamento de Matem\'atica Aplicada, UNICAMP, C.\ P.\
6065, 13083-859, Campinas, SP, Brazil}
\author{A.\ R.\ R.\ da Silva}\email{dasilva@ift.unesp.br}
\address{Instituto de F\'\i sica Te\'orica, Universidade 
Estadual Paulista,
Rua Pamplona 145, 01405-900, S\~ao Paulo, SP, Brazil}
\author{D.\ A.\ T.\ Vanzella}\email{vanzella@ifsc.usp.br}
\address{Instituto de F\'\i sica de S\~ao Carlos, Universidade de S\~ao Paulo,\\ 
Avenida Trabalhador S\~ao-carlense, 400, C.\ P.\ 369, 13560-970,
S\~ao Carlos, SP, Brazil}

\title{Can quantum mechanics fool the cosmic censor?}

\begin{abstract}

We revisit the mechanism for violating the weak cosmic-censorship 
conjecture (WCCC) by overspinning a nearly-extreme charged black hole. 
The mechanism consists of an incoming massless neutral scalar particle,
with low energy and large angular momentum, tunneling into the hole.
We investigate the effect of the large angular momentum of
the incoming particle on the background geometry and address recent 
claims that such a back-reaction would invalidate the mechanism.
We show that the large angular momentum of the  incident particle 
does {\it not} constitute an obvious impediment to the success of 
the overspinning quantum mechanism, although the induced back-reaction 
turns out to be essential to restoring the validity of the WCCC
in the classical regime. These  results seem to endorse 
the view that the ``cosmic censor'' may be oblivious to 
processes involving quantum effects.
\end{abstract}

\maketitle

For centuries our view of nature has been based on the paradigm that 
complete knowledge of the details of an {\it isolated} system at any 
particular time would determine its whole (past and future) history. 
Even the revolution unraveled by (standard) quantum mechanics was 
not enough to temper with this form of {\it determinism}, which is 
actually enforced by the unitary evolution characteristic of quantum 
theory. Notwithstanding, it is in the context of a classical theory, 
namely, general relativity (GR), that this determinism faces its
most serious threat: the {\it singularities}. ``Cosmic censors'' have 
then been postulated to oversee such unwanted objects, preventing
their existence from affecting the Universe at large, thus preserving 
the cosmic order. However, by looking at a particular simple example,
we argue that cosmic censors may be oblivious to processes involving 
quantum effects.

Singularities, which represent situations where GR itself and all 
known theories lose their predictability, are known to appear many 
times when well-posed initial conditions are evolved through Einstein
equations  (e.g., in the case of star collapse with black hole 
formation). Although it is not clear even in the classical context of 
GR whether such unpredictable objects would be able to causally 
influence ``far away'' regions, the determinism principle mentioned 
above has become so deeply rooted in the way we think about nature 
that it has motivated the formulation of the {\it weak 
cosmic-censorship conjecture} (WCCC)~\cite{Penrose69}. According to 
the WCCC, singularities should always be ``dressed up'' by event horizons 
(as in the case of black holes), thus preventing their 
``unpredictability'' from pervading the Universe. By forbidding the 
existence of ``naked'' singularities (except for a possible initial 
one), the WCCC ensures that determinism holds except possibly for 
spatially compact regions ``near'' the singularities. However, despite 
the various efforts to prove the WCCC right, its validity remains an 
elusive open question 
(see, e.g., Refs~\cite{C84,JD93,clarke,grqc,H99,penrose,C93,HHM04} and 
references therein).

Now, introducing quantum ingredients into this discussion, it is 
{\it largely} believed that a complete merging of GR with quantum 
mechanics (QM) (i.e., {\it quantum gravity}) should be able to unveil
the physical structure of singularities, making them quite benign 
{\it irrespectively} if they are naked or dressed by event horizons. 
Therefore, it is not too far-fetched to expect that the same QM might
be able to evade the WCCC, providing mechanisms 
for generating those structures. Indeed, some quantum mechanisms raising 
the possibility of formation of naked singularities have been proposed 
and discussed recently (see, e.g., Refs.~\cite{ford1,ford2,MS07,H08,RS08,H08b}). 
In particular, one such mechanism consists of a massless neutral scalar particle 
with large enough angular momentum $\sqrt{l(l+1)}$ and low enough energy 
$\omega$ being absorbed through quantum tunneling effect by a 
nearly-extreme charged black hole (with mass $M$ and charge $Q$, 
satisfying $M^2-Q^2\gtrsim 0$, and angular momentum 
$\vec J =\vec 0$)~\cite{MS07}. (We adopt units in which 
$\hbar=G=c=1$ throughout the paper.)
In the process, the black hole would acquire enough 
angular momentum (``overspin'') to become a naked singularity 
[${M}^{\prime2}-Q^2-{J}^{\prime2}/M^{\prime 2} < 0$, 
with 
${M}^{\prime}=M+\omega$ and ${J}^{\prime2}= l(l+1)$], 
thus violating the WCCC. By considering the canonical 
quantization of the low-energy sector of a free massless scalar field
in the Reissner-Nordstr\"om spacetime (see, e.g., Ref.~\cite{HMS97}), 
the probability for such a process to occur can be calculated in the 
approximation where no back-reaction is taken into account. 
The result shows that, although extremely rare, the 
overspinning mechanism is not 
forbidden as it should be in order for the WCCC to be valid.
One may wonder, however, whether or not back-reaction effects 
could come to the rescue of the WCCC.
In particular, one can investigate 
the role played by the large angular momentum of the incident 
particle on the background spacetime~\cite{H08}. Here we 
show that such a back-reaction effect is {\it not} enough to restore the
validity of the WCCC, posing no challenge to the
overspinning mechanism as long as its intrinsic quantum character
is exploited. This result contrasts with previous 
conclusions in the literature~\cite{H08} but we eventually show that both
analysis can be made consistent when properly interpreted. Interestingly 
enough, in making these proper interpretations we note that the WCCC is
restored when the classical limit of the proposed mechanism is considered.

As in Ref.~\cite{MS07}, let us begin by considering a 
nearly-extreme charged black hole with mass $M$, charge $Q$ (with 
$M^2-Q^2=\epsilon\gtrsim 0$), and angular momentum $\vec{J}=\vec{0}$, whose
line element can
be written in the form~\cite{grw}
\begin{equation}
ds^2=f(r)dt^2-f^{-1}(r)dr^2-r^2(d\theta^2+\sin^2\theta \;d\phi^2),
\label{rn}
\end{equation}
where $f(r)=(1-r_+/r)(1-r_-/r)$ and $r_\pm = M \pm \sqrt{M^2-Q^2}$.
The outer event horizon is located at $r = r_+$. 
This black hole is
assumed to be macroscopic (i.e. $M$ must be much larger than the
Planck mass $M_{\rm{P}}$) in order to guarantee the applicability of
the semiclassical gravity theory.

The mechanism proposed in Ref.~\cite{MS07}
for overspinning this black hole 
consists of sending in free massless scalar particles, {\it one 
at a time}, with low enough energy $\omega$ and angular 
momentum with large enough modulus $\sqrt{l(l+1)}$ 
($l \in {\mathbb N}$) and projection 
$m \in \{-l,-l+1,...,l-1,l\}$ in an arbitrarily given direction
(as seen from static observers 
at infinity). 
Each such particle is governed by the normal mode 
$u_{\leftarrow \omega l m}$ which is
the orthonormalized 
(according to the Klein-Gordon inner product~\cite{birrel})
solution of the 
usual Klein-Gordon equation $\nabla^{\mu}\nabla_{\mu} 
u_{\leftarrow \omega l m}=0$
subject to the condition that it is purely incoming at
the past null infinity ${\cal{I}}^-$. Writing $u_{\leftarrow \omega l m}$
in the form
\begin{equation}
u_{\leftarrow \omega l m}(t,r,\theta,\phi)=
\sqrt{\frac{\omega}{\pi}}\frac{\psi_{\leftarrow \omega l}(r)}{r} Y_{l
m}(\theta,\phi)e^{-i\omega t},
\label{ansatz}
\end{equation}
with $Y_{l m}(\theta,\phi)$ being the usual spherical-harmonic functions, 
$\psi_{\leftarrow \omega l}$ satisfy
\begin{equation}
\left[-f(r)\frac{d}{dr}\left(f(r)\frac{d}{dr}\right)+ V_{\rm
eff}(r)\right]\psi_{\leftarrow \omega l}(r)= \omega^2\psi_{\leftarrow
\omega l}(r).
\label{dre1}
\end{equation}
Here
\begin{equation}
V_{\rm eff}(r) =
f(r)\left[{l(l+1)}/{r^2} + {2M}/{r^3} - {2Q^2}/{r^4}\right]
\label{esp}
\end{equation}
is the effective scattering potential. Obviously, Eq.~(\ref{dre1}) possesses 
two independent solutions for given $(\omega, l)$ associated with modes 
(i) purely incoming from the past null infinity ${\cal{I}}^-$ and 
(ii) purely outgoing from the white-hole horizon ${\cal{H}}^-$. Here 
we  are only interested in modes (i) (labeled by the subscript
$\leftarrow$). Due to the existence of the effective scattering potential, 
low-energy ($\omega \approx 0$) incoming particles are mostly reflected back to
infinity; the {\it{few}} particles which enter
the hole must quantum-mechanically {\it tunnel} into it.

Since we are interested here only in particles coming from infinity
with low energy, we will only write the leading term in the $\omega$ 
expansion for $\psi_{\leftarrow \omega l}(r)$:
\begin{equation}
\psi_{\leftarrow \omega l}(r) \approx C_{\omega l}
\frac{(2l)!\;\bar{r}^{\;l+1}}{(l!)^2(\bar{r}_+ -\bar{r}_-)^l},
\label{lpap}
\end{equation}
where $\bar{r} \equiv r/2M$, $\bar{r}_{\pm} \equiv r_{\pm}/2M$,
and 
\begin{equation}
C_{\omega l}=
(-i)^{l+1}\frac{2^{2l+1}(l!)^{3}(\bar{r}_+-\bar{r}_-)^{l}M^{l+1}\omega^l}{(2l)!(2l+1)!}.
\label{cte}
\end{equation}
By comparing Eq.~(\ref{lpap}) with
the form that $\psi_{\leftarrow \omega l}(r)$ must exhibit near the outer
horizon,
\begin{equation}
\psi_{\leftarrow \omega l}(r) \approx (2 \omega)^{-1}{\cal
T}^{0}_{ \omega l}\; e^{-2i M\omega r^{\ast}} ; \; r^{\ast} <0,
\;|r^{\ast}|\gg 1,
\label{tail1}
\end{equation}
where
$$
r^{\ast} \equiv
\bar{r}+\frac{\bar{r}^{\;2}_{+}\ln|\bar{r}-\bar{r}_{+}|-
\bar{r}^{\;2}_{-}\ln|\bar{r}-\bar{r}_{-}|}{\bar{r}_{+}-\bar{r}_{-}}
$$
is the Regge-Wheeler radial coordinate, it follows that the probability
for the incoming particle to tunnel into the hole is, to the lowest order in $\omega$,
\begin{equation}
|{\cal T}_{ \omega l}^0|^2 = \frac{2^{2l+2}r_+^2(r_+ -
r_-)^{2l}(l!)^6 \omega^{2l+2}} {[(2l+1)! (2l)!]^2}
\label{T0}
\end{equation}
(see, e.g., Ref.~\cite{CM00} for more detail). Regardless how small such a 
probability may be, it does allow for a low-$\omega$ large-$l$ particle to be absorbed by
the nearly-extreme charged black hole which then, by symmetry
(i.e., conservation) arguments, should eventually
be characterized by a new mass $M^\prime = M+\omega$ and a new angular momentum 
satisfying $J^{\prime 2}=l(l+1)$. Therefore, a naked singularity would be formed 
provided that $M^{\prime 2}-Q^2-J^{\prime 2}/M^{\prime 2}<0$, which is true for 
a range of values of
$\omega>0$ as long as 
\begin{equation}
l(l+1) > M^2 \epsilon.
\label{trivial}
\end{equation}
If the original hole misses one single electric charge to become extreme,
then $\epsilon \approx 2 M/ \sqrt{137}$.

The fact that such large angular momenta would be necessary to challenge 
the WCCC naturally raises concern about the approximation where no back-reaction is considered.
So, in a tentative to consider the back-reaction of the particle's angular 
momentum on the spacetime, we follow Ref.~\cite{H08} and assume 
that this should be similar to the one induced by an axisymmetric ring 
of particles rotating around the black hole, which is to give some angular
velocity to the horizon generators~\cite{W74}. In the case of a 
Reissner-Nordstr\"om black hole, this angular velocity would 
be~\cite{H08}
\begin{equation}
\Omega = \frac{m_z}{M r_+^2},
\label{Omega}
\end{equation}
where $m_z$ is the ring angular momentum. 
By replacing the ring  by
the incident incoming particle, the idea is to associate  the {\it total}
angular momentum of the ring, $m_z$, with the azimuthal angular momentum
of the particle, $m$. 
Although this association is used freely
in Ref.~\cite{H08}, it deserves some comments which will be useful later.
The one-particle state characterized by the quantum numbers
$(\omega, l ,m)$ can be heuristically thought
of as an equally-weighted superposition of all the trajectories
having energy $\omega$ and angular momentum 
$\vec{L}$ satisfying both $L^{2} =
l(l+1)$ and $\vec{L} \cdot {\bf \hat{z}}=m$,
where $ {\bf \hat{z}}$
is the unit vector pointing in the direction of symmetry  of 
the incoming-particle state. Such a superposition
is obviously axisymmetric with respect to ${\bf \hat{z}}$ and has an ``averaged''
total
angular momentum given by $\langle \vec{L} \rangle = m {\bf \hat{z}} $.
Hence, the effect of such a quantum state on the {\it classical} background 
(as long as the semiclassical approximation is valid) should be 
as if the
total angular momentum were given by~$m$.

The next step, then, is to assume that the incoming particle feels the scattering
potential modified by the effective rotation given by Eq.~(\ref{Omega}).
In this case the tunneling probability becomes~\cite{H08}
\begin{eqnarray}
|{\cal T}_{ \omega l}^\Omega|^2 &=& 
\frac{2^{2l+2}r_+^2(r_+ -
r_-)^{2l}(l!)^6 \omega^{2l+1}(\omega-m\Omega)} {[(2l+1)! (2l)!]^2}
\nonumber \\
& &\times
\prod_{n=1}^l{\left[1+\left(
\frac{\omega-m\Omega}{2\pi n T_{BH}}\right)^2
\right]}
\label{HOD_probability}
\end{eqnarray}
with  $T_{BH}=(r_+-r_-)/(4\pi r_+^2)$ being the Hawking temperature of
the black hole. Note that the effect of the induced rotation 
$\Omega$ is to prevent particles with energy $ \omega \leq m \Omega$ 
from being absorbed by the hole. (In particular, the negativity of 
$|{\cal T}_{ \omega l}^\Omega|^2$  for $\omega < m \Omega$ is usually 
interpreted as the occurrence of superradiation rather than absorption.)
Nonetheless, the important point is that once {\it one} particle with 
energy  
\begin{equation}
\omega > m \Omega = m^2/(M r_+^2)
\label{oi}
\end{equation} 
tunnels in, the {\it total} 
angular momentum it transfers to the hole is determined by $l$, {\it not} 
$m$. Hence, the final state of the hole is characterized by a mass
$M^\prime = M + \omega > M+m^2/(M r_+^2)$, angular momentum $\vec{J}^\prime$ 
satisfying both $J^{\prime 2} = l(l+1)$ and $\vec{J}^\prime \cdot {\bf \hat{z}}=m$,
and the same initial charge $Q$. 
With respect to the WCCC, this implies that
the overspinning mechanism
still holds true for a range of values of $\omega>m\Omega$ provided 
$J^{\prime 2}/M^{\prime 2} > M^{\prime 2} - Q^2$, i.e.,
\begin{eqnarray}
l(l+1)>
\left(M+\frac{m^2}{M r_+^2}\right)^2
\left(\epsilon+\frac{2m^2}{r_+^2}+\frac{m^4}{M^2r_+^4}\right),
\label{newcondition}
\end{eqnarray}
where we recall that $\epsilon=M^2-Q^2$.
In particular, if the incoming particle is prepared in a state with
$m=0$ the back-reaction considered here plays no role at all since
Eq.~(\ref{oi}) becomes trivial and Eq.~(\ref{newcondition}) reduces to
Eq.~(\ref{trivial}).

These results clearly contrast with previous conclusions presented
in the literature~\cite{H08}. The point of divergence is easily identified
to be related to the angular momentum acquired by the hole when the incoming
particle tunnels
into it. The author of Ref.~\cite{H08} seems to have been carried away by the 
identification between the azimuthal angular momentum $m$
of the incoming particle
and the total angular momentum $m_z$ of a ring of particles 
[see discussion below Eq.~(\ref{Omega})].
He then concludes that
$J^{\prime 2} = m^2$, as would be natural if the hole had swallowed the ring 
of particles. Then, as the inequality~(\ref{newcondition}) is always false 
if $l(l+1)$ is replaced by $m^2$, Ref.~\cite{H08} claims that the WCCC 
has been rescued by the back-reaction effect.
However, even though the relevant angular momentum
for back-reaction purposes is given by $m$ (as previously discussed), 
once the {\it one} 
particle does get absorbed it delivers its {\it total} angular momentum
to the hole, which acts as a classical angular-momentum measuring apparatus
(provided $M\gg M_{\rm P}$).
Here lies the intrinsic quantum nature of the overspinning mechanism:
the one particle tunnels into the hole due to its wavy nature,
but it gets absorbed as a single quantum, transmitting to the hole
its energy and angular momentum.
This is similar to what happens in the double-slit 
experiment with individual particles: each particle propagates through the
double slit as a wave, but it collapses at one single spot
on the screen.

Motivated by this discussion, it 
is interesting to consider the ``classical limit'' of the
overspinning mechanism, where an {\it ensemble} of particles, all in the
same state characterized by $(\omega, l, m)$, is sent toward the 
nearly-extreme charged black hole. In this case, only a fraction 
[well approximated by Eq.~(\ref{HOD_probability})]
of the incoming particles 
would tunnel into the hole, delivering an energy $N\omega$ and a {\it total} 
angular momentum 
$\vec{J}^{\prime } = N \langle \vec L \rangle =
Nm {\bf \hat{z}} $, where $N$ (assumed to be $\gg 1$) is the number 
of absorbed particles. It is easy to note, then, that the condition for the overspinning 
mechanism to work would be the one given by inequality~(\ref{newcondition}) with 
every $m$ replaced by $Nm$ and $l(l+1)$ replaced by $ (\vec{J^\prime})^2 = N^2m^2$,
which is never satisfied; 
i.e., the overspinning mechanism would {\it fail} and the validity of the WCCC 
would be restored. This is the proper and
interesting interpretation
of the results presented in Ref.~\cite{H08}: the back-reaction induced by the 
angular momentum of the {\it ensemble} of incoming particles prevents
the violation of the WCCC. This is a {\it classical} result, in the sense that
if the ensemble of particles were able to violate the WCCC, so would a 
{\it classical} wave sent toward the hole (recall that ``tunneling'' is a 
common effect for classical waves).

In summary, we have shown that the large angular momentum of the 
incident scalar particle does not constitute an obvious impediment 
to the success of the overspinning quantum mechanism proposed in 
Ref.~\cite{MS07}. On the other hand, we have also shown,
using results of Ref.~\cite{H08}, that the back-reaction induced 
by such angular momenta does come to the rescue of the WCCC in the 
classical regime. These two results, combined, strengthen 
the view that the violation of the WCCC may be an 
intrinsic quantum process, which in turn gives support to the idea
that naked singularities might be tamed by a complete quantum gravity theory.

It is worthwhile to note at this point that back-reaction effects, 
which turns out to be the main impediment to reach a final conclusion 
about the success of the present naked singularity production mechanism, 
can be minimized (see also Ref.~\cite{RS08}). This can be achieved, e.g., 
by replacing the nearly-extreme charged hole 
by a nearly-extreme rotating one with angular momentum $L$ such that 
$M^2-L^2/M^2 = \delta \gtrsim 0$. Then a particle with modest angular 
momentum: $l(l+1) \ll L^2$, should not significantly disturb
the spacetime as it approaches the horizon and it could still overspin 
the hole  if 
$
(M+\omega)^2 - [L^2 +l(l+1)]/(M+\omega)^2 <0,
$
i.e., 
\begin{equation}
l(l+1) >  M^2 \delta + 4 M^3 \omega + {\cal O} [ M^2 \omega^2  ].
\end{equation}
We are assuming here that the azimuthal angular momentum of the particle
is null, $m=0$, and that the quantization and black hole rotation axes 
are the same. In order to avoid superradiance it is enough again to 
impose the constraint $\omega > m \Omega$, which is obviously not a 
problem for $m=0$. Clearly, a more elaborated semiclassical back-reaction 
calculation should take into account the continuous change in time of the 
scattering potential rather than assuming that the field back-reacts on 
the spacetime generating a new static scattering potential which is 
``in place" before the wavepacket tunnels through the barrier.
In spite of it, even more detailed semiclassical calculations in the 
lines above would not be enough to definitely resolve the problem. 
We note that the black hole and singularity scattering potentials are 
quite different. Eventually only a  forthcoming full quantum gravity theory 
will be able to decide whether or not there would exist some interaction 
Hamiltonian $\hat H_{\rm qg}$ 
evolving some initial state describing a particle in the spacetime of a 
black hole into a naked singularity (plus debris): 
\begin{equation}
\hat \rho_{\rm bh} \otimes |\omega lm \rangle \langle \omega lm | 
\stackrel{H_{\rm qg}}{\longrightarrow}
\hat \rho_{\rm sing} \otimes \hat \rho_{\rm debris}.
\end{equation}
Finally, we speculate about ways to preserve the generalized second law (GSL) 
if the naked singularity is revealed and raise a conjecture. An exciting idea 
would be that naked singularities and elementary particles would be 
low-energy-theory manifestations of some common quantum gravity structure, 
since all known elementary particles satisfy 
the constraint $M^2 < Q^2 + J^2/M^2$, where $M$, $Q$ and $J$ should be associated 
here with the particle's mass, electric charge and spin, respectively. This would 
explain, e.g., why elementary neutral scalar particles have never been observed 
(since in this case $Q=J=0$), and imply that the Higgs boson, if observed in 
the LHC/CERN, would be a composite scalar particle. In this scenario, a
naked singularity would decay into a myriad of elementary particles which would
carry a hopefully large enough entropy to preserve the GSL. (See Ref.~\cite{GJS06}
for a loop quantum gravity discussion on the ``quantum evaporation of naked
singularities" which seems to be in line with our present conjecture.)
In contrast to it, singularities hidden in the interior of event horizons 
would be stable because of the very spacetime structure.

\acknowledgments

All the authors would like to acknowledge financial support from 
Funda\c{c}\~ao de Amparo \`a Pesquisa do Estado de  S\~ao Paulo (FAPESP). 
G.\ M., A.\ S., and D.\ V.\ are also grateful to Conselho Nacional de 
Desenvolvimento Cient\'\i fico e Tecnol\'ogico (CNPq) for partial 
financial support.

\end{document}